\documentclass[12pt]{article}\usepackage[hyperfootnotes=false]{hyperref}
\usepackage{epsfig}
\usepackage{float}
\usepackage{empheq}
\usepackage{bbold}

\usepackage[utf8]{inputenc}
\usepackage{amsmath}

\usepackage{caption}

\usepackage{amsmath}
\usepackage{amssymb}
\usepackage{graphicx}
\newlength{\abstractwidth}
\renewcommand{\thefootnote}{\fnsymbol{footnote}}
\renewcommand{\thanks}[1]{\footnote{#1}} 
\newcommand{\starttext}{
\setcounter{footnote}{0}
\renewcommand{\thefootnote}{\arabic{footnote}}}

\newcommand{\be}{\begin{equation}}
\newcommand{\bea}{\begin{eqnarray}}
\newcommand{\eea}{\end{eqnarray}}
\newcommand{\beq}{\begin{equation}}
\newcommand{\ee}{\end{equation}}

\newcommand*\widefbox[1]{\fbox{\hspace{2em}#1\hspace{2em}}}

\def\eq{&=&}

\def\la{\langle}
\def\ra{\rangle}
\def\simleq{\; \raise0.3ex\hbox{$<$\kern-0.75em
\raise-1.1ex\hbox{$\sim$}}\; }
\def\simgeq{\; \raise0.3ex\hbox{$>$\kern-0.75em
\raise-1.1ex\hbox{$\sim$}}\; }

\def\bi{\begin{itemize}}
\def\ei{\end{itemize}}

\def\dof{degrees of freedom }

\def\CJ{{\cal{J}}}

\def\CT{{\cal{T}}}

\def\t{\tau}

\def\bsub{ \begin{subequations}
\begin{empheq}[box=\widefbox]{align}  }
\def\esub{ \end{empheq}
\end{subequations}}

\def\1{\(  \mathbb{1} \)}

 \def\lf{\left(}
    \def\rg{\right)}

  \def\bn{\bigskip \noindent}

 \def\bm{\begin{bmatrix}}
 \def\em{\end{bmatrix}}

    \def\J-{\CJ^{-1}}

\makeatletter
\g@addto@macro\normalsize{%
  \setlength\abovedisplayskip{10pt}
  \setlength\belowdisplayskip{20pt}
  \setlength\abovedisplayshortskip{10pt}
  \setlength\belowdisplayshortskip{20pt}
}
\makeatother

\usepackage{color}

\begin{document}


  
\begin{titlepage}

 \rightline{}
\bigskip
\bigskip\bigskip\bigskip\bigskip
\bigskip

\centerline{\Large \bf {Infinite Temperature's Not So Hot    }} 

\bn

\bigskip
\begin{center}
	\bf  Henry Lin$^{a}$ and  Leonard Susskind$^{b}$  \rm

\bigskip

$^{a}$ Jadwin Hall, 
Princeton University,
Princeton, NJ 08540, USA\\

$^{a}$ Institute for Advanced Study,
Princeton, NJ 08540, USA 
\\

$^{b}$ Stanford Institute for Theoretical Physics and Department of Physics, \\
Stanford University,
Stanford, CA 94305-4060, USA \\

\end{center}

\bn

\begin{abstract}

It has been argued that the entanglement spectrum of a static patch of de Sitter space must be flat, or what is equivalent, the  temperature parameter in the Boltzmann distribution must be infinite. This seems absurd: quantum fields in de Sitter space have thermal behavior with a finite  temperature proportional to the inverse radius of the horizon. The resolution of this puzzle is that the behavior of some quantum systems 
can be characterized by a  temperature-like quantity which  remains finite as the temperature goes to infinity.
For want of a better term we  have called this quantity tomperature. In this paper we will explain how tomperature resolves the puzzle in a proposed  toy model of de Sitter holography---the double-scaled limit of SYK theory.

\end{abstract}

\end{titlepage}

\starttext \baselineskip=17.63pt \setcounter{footnote}{0}

\tableofcontents

\Large

\section{The Temperature of De Sitter Space is Infinite}
The probability for a fluctuation to take  place in de Sitter space is given by $e^{-\Delta S}$, where the entropy deficit $\Delta S$ is the decrease in entropy accompanying the fluctuation \cite{Susskind:2021omt}. For this to make sense the entropy deficit must always be positive. It follows that the entropy of the de Sitter vacuum must have  the maximum possible value.    Recently Chandrasekharan, Penington, and Witten have observed an important consequence:  the entanglement spectrum of a static patch must be flat: equivalently the density matrix of the static patch must be maximally mixed. To put it another way, the formal temperature  in the Boltzmann distribution must 
 be infinite\footnote{ Chandrasekharan, Penington, and Witten express this in terms of Von Neumann algebras: the operator algebra  of the static patch should be of type II.  The same physical conclusions were reached long ago by Banks     \cite{Banks:2003cg} and Fischler \cite{Fischler}, and more recently by
 Dong, Silverstein, and Torroba \cite{Dong:2018cuv} on the basis of  different arguments.}.


\subsection{Global and Proper Temperature}
Consider quantum field theory in a background de Sitter space. The metric in static coordinates is given by,
\be 
ds^2 = l^2 \lf -(1-r^2) dt^2   +  (1-r^2)^{-1} dr^2 +r^2 d\Omega^2  \rg
\ee
where $l$ is the characteristic de Sitter length scale.
We assume that there is a Hamiltonian generating time-translations,
\be 
H =l^{-1} \frac{\partial}{\partial t}.
\ee
The factor $l^{-1}$ is to give $H$ the units of energy.

The definition of global temperature is through the usual Boltzmann distribution,
$$ \frac{1}{Z}   e^{-H/T} =   \frac{1}{Z} e^{-\beta H}.$$ For quantum field theory in a de Sitter background the global temperature is given by,
\be 
T_{dS}=\frac{1}{2\pi l}
\ee

One can also consider the local proper temperature---the temperature  that would be registered by a static thermometer at spatial coordinate $r$. One can also think of it as the local Hawking temperature of radiation emitted from the horizon. The proper temperature
is related to the global temperature by a red-shift factor,
\be 
T_{proper} = \sqrt{1-r^2} \  T_{dS}.
\label{glob loc}
\ee
At the center of the static patch  where $r=0$, the proper temperature is the same as the global temperature.

However, according to \cite{Witten}      when dynamical gravity is ``turned on"  the  Boltzmann distribution must become flat, and the global temperature  infinite.
But if the global temperature is infinite then by \eqref{glob loc}  the proper temperature must  also be infinite---everywhere.  

Off hand this sounds nonsensical. If the proper temperature  were infinite we would  be burned to a crisp by the radiation from the cosmic horizon.

\subsection{Tomperature}

We will resolve this  puzzle by showing  that systems of discrete \dof \ (qubits for example) at infinite temperature can behave thermally with an effective temperature which remains finite as $T\to \infty.$ The effective temperature, denoted $\CT$ will be called tomperature  \cite{Lin:2019kpf}. We will define tomperature, and then show that in a toy model of de Sitter space  field correlations have thermal form, with the effective temperature being the tomperature. 

The example we will concentrate on is the infinite temperature limit of double-scaled  SYK.

\subsection*{NOTE}
The double-scaled SYK theory (DSSYK)  \cite{Cotler:2016fpe}\cite{Berkooz:2018jqr} is usually defined   as the limit 
$$
N \to \infty,  \ \ \
q \sim  N^{1/2}
$$ 
In this
paper, as in \cite{Susskind:2021esx} and  \cite{Susskind:2022dfz}   we will mean something a bit more general; namely the limit 
\be
N \to \infty, \ \ \ \     q \sim  N^{p}    
\label{defDSSYK}
\ee
with $$0<p<\frac{1}{2}$$.
 
The value of $p$ will not be important in this paper but it may be constrained when $1/N$ corrections are considered.

\section{Tomperature in DSSYK}
\subsection{Conventional SYK Scaling }
The SYK model  \cite{Maldacena:2016hyu}  is defined by the following equations,
\be 
H =i^{q/2} \sum_{1\leq i_1<i_2...i_q \leq N} j_{i_1i_2....i_q} \psi_{i_1}\psi_{i_2}...\psi_{i_q}
\label{SYKH1}
\ee
\be 
\{ \psi_a, \psi_b  \} = \delta_{ab}
\label{acomm1}
\ee
\be 
\la  j^2  \ra = \frac{2^{q-1}(q-1)!}{q N^{q-1}} \CJ^2
\label{Var1}
\ee

By  rescaling the fermion \dof 
$$\chi_i = \sqrt{2} \psi_i.$$  the annoying factors of $2^q$ in \eqref{Var1} can be removed.

\subsection{Scaling for DSSYK}

To keep the Hamiltonian finite in  the double-scaled limit 
 we must rescale $H$ by multiplying it by $q.$ With these changes equations 
\eqref{SYKH1}\eqref{acomm1}\eqref{Var1} become,
\be 
H =i^{q/2} \sum_{1\leq i_1<i_2...i_q \leq N} j_{i_1i_2....i_q} \chi_{i_1}\chi_{i_2}...\chi_{i_q}
\label{SYKH}
\ee
\be 
\{ \chi_a, \chi_b  \} = 2\delta_{ab}
\label{acomm}
\ee
\be 
\la  j^2  \ra = \frac{q!}{ N^{q-1}} \CJ^2
\label{Var}
\ee

The rescaling of the Hamiltonian by a factor $q$ is equivalent to rescaling time by the inverse factor. In using results from other papers we will have to take this  re-scaling of time into account. In particular when comparing with \cite{Roberts:2018mnp} wherever $t$ appears it will be  replaced by  with $qt$ (see for example section \ref{5.1}).

In the standard SYK theory with fixed $q$ the rescaling of $H$ would  trivially rescale  the units of time and energy. But in DSSYK the re-scaling is essential for the Hamiltonian (and other important quantities)   to remain finite as the double-scaled limit is taken. This theme will recur throughout the paper.

\section{Infinite Temperature}
\subsection{A System of Particles}
Let's begin with an ordinary gas of $N$ weakly coupled particles. The temperature is defined by the parameter in the Boltzmann distribution,
\bea 
\rho \eq \frac{1}{Z} e^{-\beta H} \cr \cr
Z \eq Tr \ e^{-\beta H} \cr \cr
T \eq 1/\beta
\label{defB}
\eea

At high temperature the system behaves classically. The energy per particle  goes to infinity,
\be 
\frac{E}{N} \to \frac{3}{2}T \to \infty.
\label{class}
\ee
Time scales, such as the mean time between collisions, the thermalization time, diffusion time, and scrambling time all go to zero. The time-scale for the decay of correlation functions also goes to zero.

\subsection{A System of Qubits}
Let's compare this with a system of $N/2$  qubits interacting through $q$-local all-to-all couplings, as exemplified by the SYK system. 
For such systems the temperature is still defined by \eqref{defB} but the energy per qubit remains finite at $T\to \infty.$ The time-scales all go to finite limits and correlation functions decay at finite rates. The question we address is whether there is a single finite temperature-like quantity which characterizes the energetics and time scales. Our answer is yes. For want of a better name we call that quantity ``tomperature" and denote it by $\CT.$ We claim that the quantity which is usually identified with the temperature of de Sitter space is actually the tomperature.

\subsection{What We Are Not Saying}
 To be clear about what we are saying we first  explain what we are not saying. Consider a box of particles $\bf B$ in contact with a  system of qubits $\bf H,$ located at the walls of the box:  the entire system is assumed to be in equilibrium. For such a system the temperature of the two subsystems must be the same. If $\bf H$ is at infinite temperature $\bf B$ must also be at infinite temperature. Anyone who comes in contact with $\bf B$ will get burned. There is no sense in which $\bf B$ has an  finite effective temperature $\CT.$
 
 How is this different from de Sitter space with its horizon  at infinite temperature, and its bulk at an effective low tomperature?  The answer is that in the first case $\bf B$ and $\bf H$ are independent subsystems described by a product Hilbert space, and commuting degrees of freedom. In the second case the horizon system is all that there is. The bulk is not a second subsystem; it's a holographic construct made of the horizon degrees of freedom. As we will see, a bulk can emerge  at finite effective tomperature, from  a hologram at infinite temperature.

\section{Tomperature in SYK}
\subsection{Definition of Tomperature}

The definition of tomperature is inspired by the analogous definition of temperature. From the first law,
$$
T = \frac{dE}{dS}.
$$
In other words the temperature is the  \it change in energy when the entropy is changed by one unit. \rm \  In this definition it is assumed that the parameters of the system, namely the number of degrees of freedom $N$, and the couplings $j$ are held fixed. Obviously, under these restrictions, at infinite temperature $dE/dS$ is infinite. In defining tomperature we will consider a different way of varying the entropy that leads to a finite result for Tomperature.

At infinite temperature the entropy is simple the half the number of fermionic coordinates (each coordinate counts as half a qubit),
\be 
S=N/2.
\label{S=N/2}
\ee

\bn
\bf Definition:\rm

\bn
\it Tomperature is  the change in energy if we remove one qubit, i.e.,  two fermionic degrees of freedom,  while keeping fixed the couplings  involving all   other fermions. \rm
\bn

\subsection{Calculation of Tomperature}

We will now calculate the tomperature.
Naively all we have to do is to compute the energy per fermion (relative to the ground state)   in the infinite temperature ensemble and multiply by $2.$   For $p<1/2$ the energy per fermion is given by,
\be 
\frac{E}{N} = \frac{\CJ}{q}
\label{E/N=J/q}
\ee
 (see \cite{Maldacena:2016hyu},  equation 2.32) so that removing two fermions would give an energy  change $2\CJ/q.$ But this calculation assumes that the normalization of the couplings changes according to \eqref{Var} when $N$ goes to $N-2.$ The right rule is that those couplings---all the ones not involving the deleted fermions, should {\it not} change when the qubit is deleted. So we need to correct the new energy by a multiplicative factor $ \left(  \frac{N-2}{N}  \right)^{\frac{q-1}{2}}$.

Taking this into account,
\be  
 \Delta E =\frac{N\CJ}{q} -\frac{(N-2)\CJ}{q}\lf  \frac{N-2}{N}     \rg^{\frac{q-1}{2}}
 \label{calculation}
\ee
which for large $N$ and $q$ is given by,
$$
 \Delta E \approx 2{\CJ}.
$$
Thus the tomperature is,
\be 
\CT = 2 \CJ.
\label{result}
\ee
Remarkably it depends only on $\CJ.$ 

In the proposed correspondence with de Sitter space the energy scale $\CJ$ is identified with the Hubble scale $l^{-1}.$ The tomperature  is both the energy cost of removing a fermion,  and the energy of a single Hawking quantum with a wavelengh  $\sim l$.

\subsection{Hawking Temperature Equals Tomperature}

Earlier we explained that  the claim of infinite static-patch temperature   was motivated by the formula for the probability for fluctuations:
\be 
{\rm{Prob}} = e^{-\Delta S}.
\label{Prob}
\ee
Let us consider an example---the probability that a single qubit becomes disconnected from the horizon degrees of freedom. This is exactly the situation that was envisioned in the definition of tomperature. Thus we may write,
\be 
\Delta E = \CT \Delta S
\label{DE=CTDS}
\ee
where $\Delta E$ is the change in the energy of the horizon when a qubit is emitted. It is also the energy carried off by the qubit. Combining \eqref{Prob} and \eqref{DE=CTDS}, the probability for the emission of a qubit from the horizon is,
\be 
{\rm{Prob}} = e^{- \frac{\Delta E}{\CT}}.
\ee	
This is  what one expects for the emission of Hawking radiation--if one identifies the Hawking temperature with the tomperature.
\bn

In the proposed correspondence with de Sitter space the energy scale $\CJ$ (and therefore $\CT$) is identified with the Hubble scale $l^{-1}.$  Thus the tomperature  is both the energy cost of removing a fermion,  and the energy of a single Hawking quantum with a wavelengh  $\sim l$.

\section{Correlation Functions}
\subsection{The Two-Point Function from SYK}\label{5.1}
Let us consider the two-point function $G =\la \chi(0) \chi(t) \ra$  in SYK. At large $q$ and  infinite temperature it was computed in \cite{Roberts:2018mnp}. In quoting the result we must remember to take account of the re-scaling of time by a factor of $q.$ With that taken into account the result of \cite{Roberts:2018mnp} is,
\be 
G(t) = \lf  \frac{1}{\cosh{(q\CJ t)}}    \rg^{2/q}.
\label{G}
\ee
In the limit of large $q,$ $G$ tends to a simple form,
\be 
G(t) = e^{-2\CJ |t|}
\label{Gexp}
\ee
or from \eqref{result},
\be 
G(t) = e^{- \CT |t|}
\label{Gexptomp}
\ee

It was not obvious that $G$ should    tend to a $q$-independent function.
The fact that it does so  is an essential requirement for a correspondence between DSSYK and de Sitter space. The behavior of correlation functions in the vicinity of a horizon is  a manifestation of the existence of quasi normal modes. The result \eqref{Gexptomp} is characteristic of  the exponential decay of  quasi normal modes  with the rate being proportional to the Hawking temperature\footnote{We could try to apply the same logic to the fixed-$q$ case at infinite temperature as a model for a far-from-extremal black hole. In that case we find a mismatch between the tomperature and the behavior of correlation functions. This may not be surprising since the analysis of the operator algebras of black holes does not lead to a flat spectrum. }.

\subsection{The Bulk Two-Point Function }

We now consider a typical two-point function in  the bulk, i.e., the portion of the static patch between the stretched horizon and $r=0.$ The holography of Sitter space assumes that the holographic \dof \ live on the stretched horizon. Therefore to compare with \eqref{Gexptomp} we will calculate the field-field correlation function between two points on the stretched horizon but in the bulk theory.

\begin{figure}[H]
\begin{center}
\includegraphics[scale=.55]{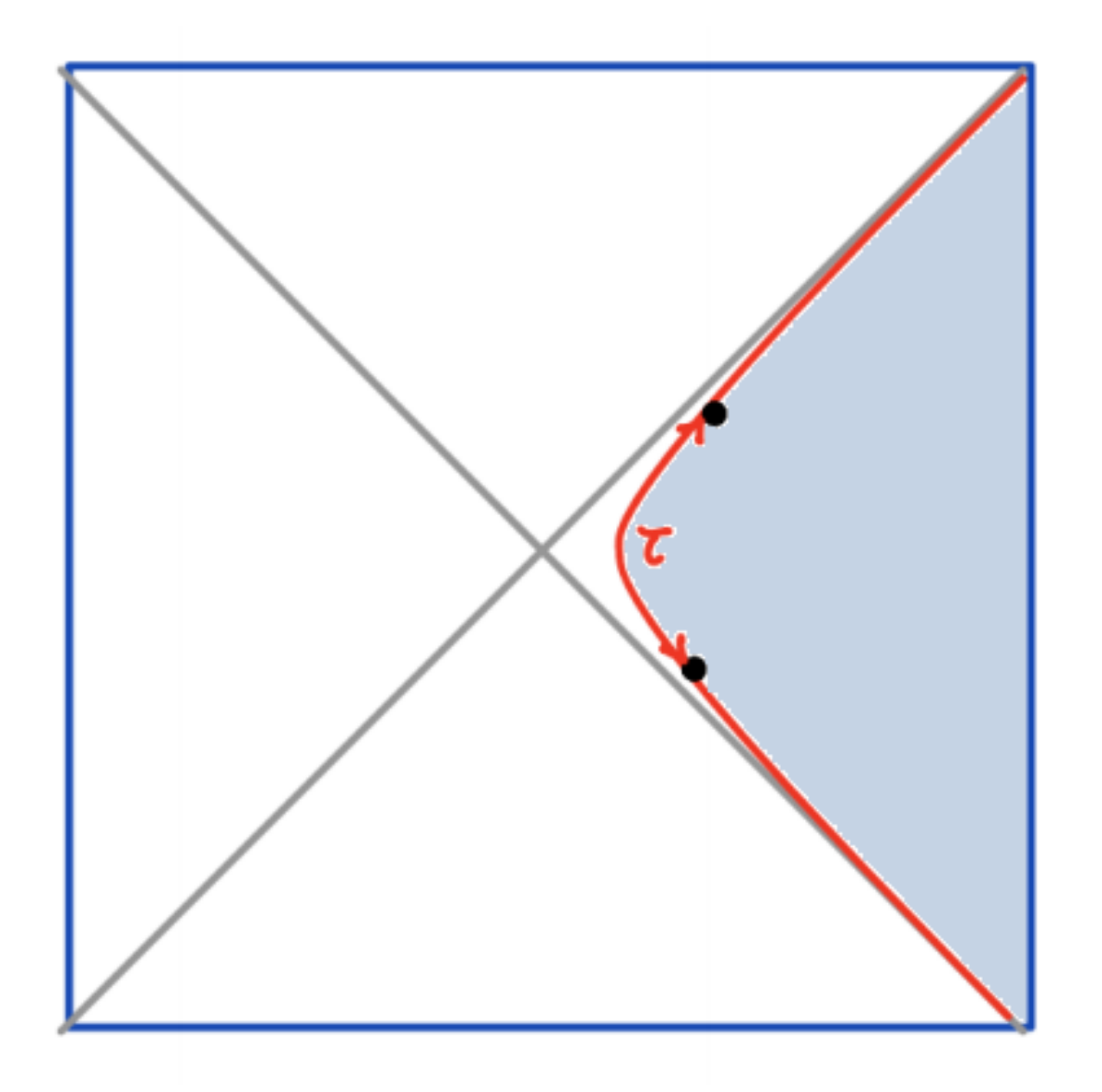}
\caption{Two points on the stretched horizon separated by a boost angle $\tau$}
In the holographic description the dynamics is described by degrees of freedom and a Hamiltonian which live on the stretched horizon. The shaded grey region has to be reconstructed from the holographic \dof.
\label{tau2}
\end{center}
\end{figure}

In figure \ref{tau2} the de Sitter Penrose diagram is shown with the stretched horizon indicated in red. The stretched horizon is a surface at a distance $\rho$ from the true horizon. The holographic SYK \dof \ and  Hamiltonian may be visualized as living on the stretched horizon. The function \eqref{Gexptomp} was computed by studying the  of evolution of the SYK system with no reference to the existence of a bulk. 

\begin{figure}[H]
\begin{center}
\includegraphics[scale=.4]{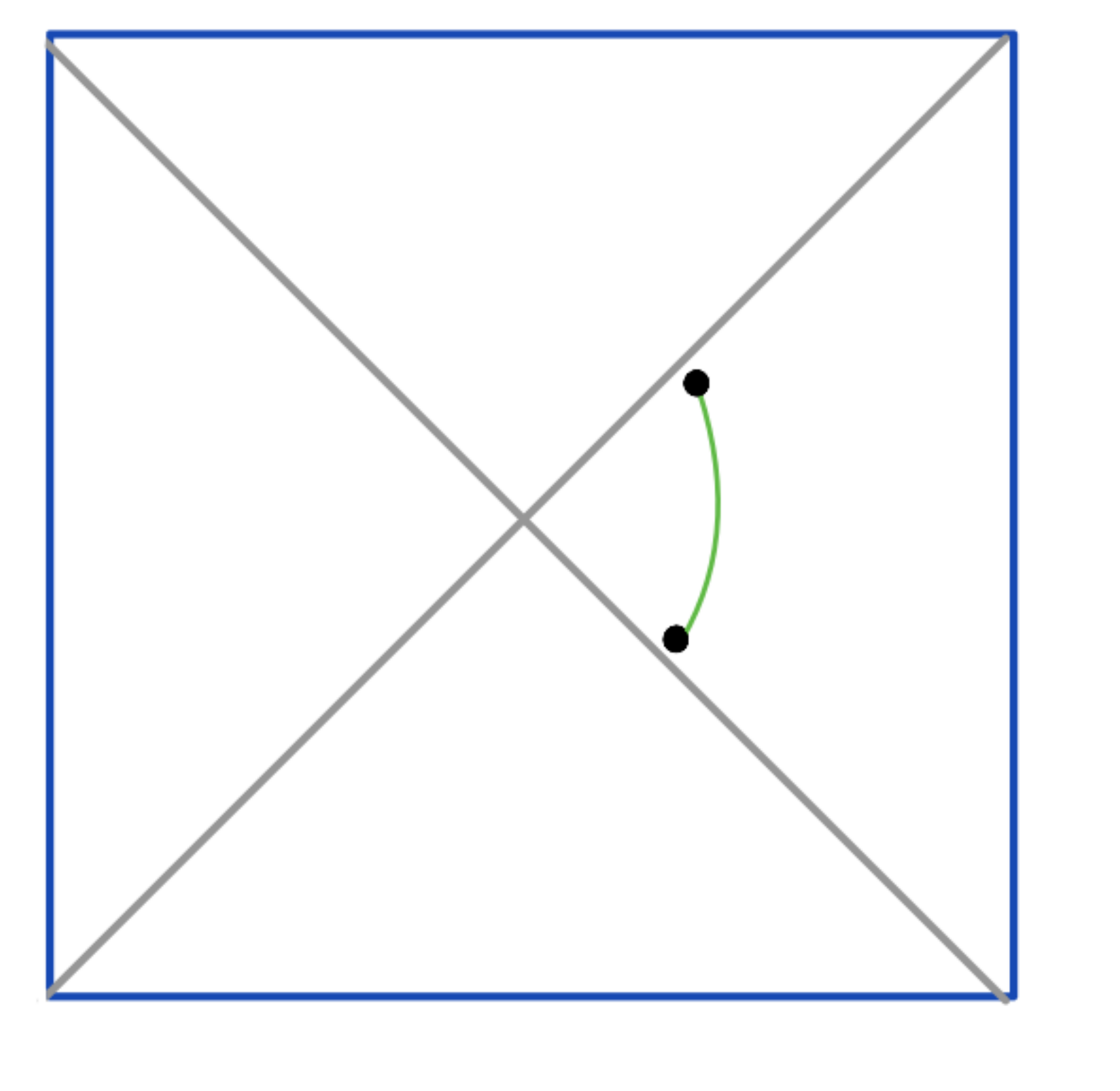}
\caption{The horizon-horizon correlation probes the bulk of the static patch.}
In the bulk description the two-point function would be dominated by paths that go through the bulk of the static patch.
\label{particle}
\end{center}
\end{figure}

The shaded grey region in figure \ref{tau2} is the portion of the static patch which must be reconstructed from the hologram. If such a reconstruction is actually possible then it 
 must also be possible to understand the  correlation function $G(\tau)$  in terms of a signal
propagating through the bulk.  Figure \ref{particle} shows a path through the bulk connecting the two horizon points. 

To compute  bulk propagators between the two horizon points for $t<l$  it is sufficient to use the Rindler approximation to the geometry.
In the Rindler approximation the proper geodesic time between the two points is,
\be 
d = \rho \cosh{\t}
\label{d}
\ee

As an example consider a bulk field in the static patch with bulk dimension $\Delta =1.$ The correlation function is,
\bea
C \eq \lf \frac{1}{d} \rg^{2} \cr \cr
&\sim& \frac{1}{\rho^{2}\cosh^2{\t}} \cr \cr
\eq \frac{1}{\rho^{2}}e^{-2| t/l|}.
\label{C formula1}
\eea
The prefactor $1/\rho^2$ can be removed by re-scaling the bulk field in which case,
\be 
C=e^{-2|t/l|}
\label{C formula}
\ee
Using  $\CT=1/l$ (see  \ref{l=1/J}), the similarity of  \eqref{C formula} and \eqref{Gexptomp} is obvious.

To emphasize the point,  the  correlator  in \eqref{C formula} describes the propagation of a signal through the bulk. It can be visualized as a sum over paths that jump out from the past horizon, pass through the static patch, and then fall back into the future horizon. The fact that it qualitatively agrees with 
 \eqref{Gexptomp}  indicates that DSSYK correlation functions know about, the bulk geometry of the static patch.

\section{Operator Growth}
We will briefly review the results of \cite{Susskind:2021esx}\cite{Susskind:2022dfz}. 
The operator growth---aka scrambling---behavior of SYK at infinite temperature can be understood in terms of the epidemic model. For the $q$-local version the epidemic model for operator growth is described by the equation,
\be 
P(\t +\epsilon) = P(\t)+ \epsilon[1-P(\t)] \ [   1-\big(  1-P(\t) \big)^{q-1}]
\label{recursion}
\ee
where $P$ is the probability that a given qubit is infected, and $\epsilon$ is the probability of transmission at an encounter. The time-variable $\tau$ is the so-called circuit time.

By taking the limit $\epsilon\to 0$ the equation can be converted to a differential equation and solved. The solution is,
 \be  
P(\tau) = 1 - \lf1+ \frac{q}{N} e^{(q-1)\tau}\rg^{\frac{-1}{q-1}}
\label{master}
\ee

For fixed $q$ and small $\tau,$ 
\be 
P(\tau) = e^{((q-1)\t}
\label{P(smalltau)}
\ee
for large $q$ this early exponential growth is very fast. The reason is obvious: at each encounter an infected qubit infects $(q-1)$ other qubits. But in the  limit we will be interested in, the exponential growth shuts down after a period which shrinks to zero as $q$ grows    \cite{Susskind:2021esx}\cite{Susskind:2022dfz}.

Now let us compare this with the result of \cite{Roberts:2018mnp} for scrambling. The initial exponential growth predicted in that paper is

\be 
P= e^{\CJ q t}
\label{P=expJqt}
\ee
where the factor of $q$ (which does not appear in \cite{Roberts:2018mnp}) is once again due to the re-scaling of time.  By comparing \eqref{P(smalltau)} with \eqref{P=expJqt},
in the large $q$ limit we find,
\be 
\t = \CJ t.
\label{tau=CJt}
\ee

Going back to \eqref{P=expJqt} we see that the exponential growth of $P(t)$ is extremely fast and diverges in the double-scaled limit. 
 But as explained in \cite{Susskind:2021esx}\cite{Susskind:2022dfz}, as $q$ grows, the time-interval over which \eqref{P=expJqt} is correct shrinks to zero. In the double-scaled limit this interval disappears altogether and and 
 \eqref{master} uniformly tends to
$$ 
P(\t)=1-e^{-\tau}
$$
or
\bea
1-P(t) \eq  e^{-\CJ t} \cr \cr
\eq e^{-\CT t}
\label{1-P=exp-Jt}
\eea

As in the case of two-point correlation functions, operator growth has a well defined behavior in the double-scaled limit. In both cases the  
 tomperature $\CT$ replaces the conventional Hawking  temperature  in correlation functions and decay rates, as well as in quantities like the energy per degree of freedom.

\section{Comparison with De Sitter}

Equations \eqref{result} \eqref{Gexptomp} and \eqref{1-P=exp-Jt} illustrate  the central point of this paper; that  quantities of physical significance in de Sitter space have good limits in the infinite temperature double-scaled limit:

\begin{enumerate}
\item
Equation \eqref{result}  shows that, although the temperature diverges,  the tomperature is finite in the limit. This was not obvious; it might have diverged or tended to zero  as $q\to\infty $.
\item
Equation \eqref{Gexptomp} shows that the two-point function and the decay constant for quasi normal modes have  good limits; something which was also not obvious. Moreover the decay constant is the tomperature which parallels the fact that in the semiclassical theory the decay constants for quasi normal modes are, to within numerical constants, the conventional de sitter temperature.
\item
Finally \eqref{1-P=exp-Jt} shows that the functional form for hyperfast scrambling  \cite{Susskind:2021esx}\cite{Susskind:2022dfz}, has a  good limit.
\end{enumerate}

But not all quantities have  limits; for example in \eqref{P=expJqt} the Lyapunov exponent for operator growth is $\CJ q$, which diverges as $q\to \infty.$ However that exponent has no meaning in de Sitter space, or for that matter in DSSYK. As explained in \cite{Susskind:2021esx}\cite{Susskind:2022dfz} 	a theory in which the holographic \dof \ are at the horizon is not a fast scrambler--it is a hyperfast scrambler. A finite period of Lyapunov growth would be incompatible with this, but as \cite{Susskind:2021esx}\cite{Susskind:2022dfz} 
 show, the period of
Lyapunov behavior  shrinks to zero as $q\to \infty.$ The final result is a hyperfast behavior with a perfectly finite limit \eqref{1-P=exp-Jt}.

Another quantity that doesn't have a finite (non-zero) limit is the energy per qubit \eqref{E/N=J/q}. Unlike the tomperature, which has an interpretation as the Hawking temperature, this quantity has no  meaning in semi-classical de Sitter space. 

These facts support the interpretation of DSSYK  as  a holographic model in which the degrees of freedom lie at the horizon, not on some distant boundary. A natural candidate for this kind of holography is de Sitter space.

\bn

There is a single dimensional parameter in the classical de Sitter metric; namely the horizon radius $l.$ Similarly in the double-scaled limit of SYK at infinite temperature there is a single dimensional parameter, $\CJ.$ In the correspondence between the two theories these parameters must be related,
\be 
l = 1/\CJ.
\label{l=1/J}
\ee
We see that we can also express this as a relation between the de Sitter radius and the tomperature.
\be 
\CT=1/l.
\label{tomp=1/l}
\ee


\section{Summary}

To summarize: Explicit calculations \cite{Dong:2018cuv}, as well as general principles \cite{Witten}, require the entanglement spectrum of a static patch to be flat, or equivalently the temperature to be infinite. Nevertheless we require that field correlations in de Sitter space behave thermally with effective temperature $1/(2\pi l).$ One might have thought that all the degrees of freedom would  come to equilibrium at the same temperature, but we have seen by the specific example of DSSYK that the finiteness of the effective  temperature  and the infinite value of the mathematical temperature coexist quite comfortably; the effective temperature being the tomperature, defined by an analog of the first law,
\be  
\delta E = \CT \delta S.
\ee

Remarkably the physically relevant quantities in de Sitter space such as the Hawking temperature, correlation functions, QNM decay rates remain finite in the infinite temperature double-scaled limit, and are given in terms of the tomperature. Other quantities that have no obvious meaning for de Sitter space diverge or vanish. 

On the role of the parameter $p$ in \eqref{defDSSYK}:
so far $p$ has not appeared in our analysis except in so far as it tells us to take $q\to \infty.$ Any value of $p$ in the range  
$0<p<\frac{1}{2}$ will give the same results.  We expect that this will change  when $\frac{1}{N}$ corrections are taken into account. We will leave this for another time.

\bn

\subsection*{A Caveat:}

  At best DSSYK is a toy model of de Sitter holography. Like its usual low-temperature AdS(2) counterpart it lacks the ingredients that are needed for locality on scales smaller than ${\CJ}^{-1}$. Roughly speaking it is analogous to string theory in which the Planck scale is microscopic but the string
scale is comparable to the cosmological  scale.

How is it that the cosmological scale, measured in Planck units, is stable without fine-tuning? The DSSYK model seems to be an example of ``set it and forget it.” Why is there no need for fine-tuning? 
This question is not unique to de Sitter space; the same issue comes up in the conventional SYK theory, except that the cosmological constant is negative.

The answer is not that SYK has found a way around the fine-tuning argument. It’s just that the cutoff scale (the string scale) is the same as the cosmological scale, namely $l$. A theory that is so non-local would not generate significant ``radiative corrections.”

\section*{Acknowledgements}
We have benefited from discussions with Geoff Penington, Edward Witten, Ying Zhao, and Douglas Stanford.

\end{document}